\documentclass[sigconf]{acmart}

\usepackage{multirow}

\AtBeginDocument{
  }

\copyrightyear{2025}
\acmYear{2025}
\setcopyright{acmlicensed}
\acmConference[SIGIR '25]{Proceedings of the 48th International ACM SIGIR Conference on Research and Development in Information Retrieval}{July 13--18, 2025}{Padua, Italy}
\acmBooktitle{Proceedings of the 48th International ACM SIGIR Conference on Research and Development in Information Retrieval (SIGIR '25), July 13--18, 2025, Padua, Italy}
\acmDOI{10.1145/3726302.3730260}
\acmISBN{979-8-4007-1592-1/2025/07}

\begin{document}

\title{Unbiased Collaborative Filtering with Fair Sampling}

\author{Jiahao Liu}
\orcid{0000-0002-5654-5902}
\affiliation{
  \institution{Fudan University}
  \city{Shanghai}
  \country{China}}
\email{jiahaoliu21@m.fudan.edu.cn}

\author{Dongsheng Li}
\orcid{0000-0003-3103-8442}
\affiliation{
  \institution{Microsoft Research Asia}
  \city{Shanghai}
  \country{China}}
\email{dongshengli@fudan.edu.cn}

\author{Hansu Gu}
\orcid{0000-0002-1426-3210}
\affiliation{
  \institution{Independent}
  \city{Seattle}
  \country{United States}}
\email{hansug@acm.org}

\author{Peng Zhang}
\orcid{0000-0002-9109-4625}
\authornote{Corresponding author.}
\author{Tun Lu}
\orcid{0000-0002-6633-4826}
\authornotemark[1]
\affiliation{
  \institution{Fudan University}
  \city{Shanghai}
  \country{China}}
\email{zhangpeng\_@fudan.edu.cn}
\email{lutun@fudan.edu.cn}


\author{Li Shang}
\orcid{0000-0003-3944-7531}
\author{Ning Gu}
\orcid{0000-0002-2915-974X}
\affiliation{
  \institution{Fudan University}
  \city{Shanghai}
  \country{China}}
\email{lishang@fudan.edu.cn}
\email{ninggu@fudan.edu.cn}


\renewcommand{\shortauthors}{Jiahao Liu et al.}

\begin{abstract}
Recommender systems leverage extensive user interaction data to model preferences; however, directly modeling these data may introduce biases that disproportionately favor popular items.
In this paper, we demonstrate that popularity bias arises from the influence of propensity factors during training.
Building on this insight, we propose a fair sampling (FS) method that ensures each user and each item has an equal likelihood of being selected as both positive and negative instances, thereby mitigating the influence of propensity factors.
The proposed FS method does not require estimating propensity scores, thus avoiding the risk of failing to fully eliminate popularity bias caused by estimation inaccuracies.
Comprehensive experiments demonstrate that the proposed FS method achieves state-of-the-art performance in both point-wise and pair-wise recommendation tasks.
The code implementation is available at \url{https://github.com/jhliu0807/Fair-Sampling}.
\end{abstract}

\ccsdesc[500]{Information systems~Recommender systems}

\keywords{unbiased ranking, collaborative filtering, implicit feedback}

\maketitle

\section{Introduction}
Recommender systems gather vast amounts of data on user behavior, especially implicit feedback from user-item interactions, to model user preferences~\cite{xia2022fire,liu2023personalized,liu2023autoseqrec,liu2022parameter,liu2023recommendation,liu2023triple,liu2024filtering}.
However, user interactions are shaped not only by intrinsic user preferences but also by external factors, such as item popularity~\cite{yang2018unbiased}.
Directly modeling observed user interactions may lead to predictions that are overly skewed toward popular items~\cite{wan2022cross}.
The Matthew effect exacerbates this by making long-tail items less likely to be recommended, thus diminishing the novelty, diversity, and fairness of recommendations~\cite{klimashevskaia2024survey}.

Several methods based on inverse propensity scores (IPS)~\cite{saito2020unbiased,saito2020unbiased2} have been proposed to mitigate popularity bias by weighting samples during model training.
However, estimating propensity scores in practical applications can be highly challenging.
These methods often rely on overly simplified assumptions---for instance, assuming that the propensity score depends solely on item popularity---which can lead to high variance~\cite{ren2023unbiased}.
Although variance reduction techniques, such as clipping the scores, can be applied, these adjustments compromise unbiasedness.
Furthermore, the difficulty in accurately estimating propensity scores prevents IPS-based methods from effectively eliminating popularity bias.

In this paper, we first analyze the optimization objectives of ideal and classical loss functions, identifying that the bias in classical loss functions stems from the inclusion of propensity factors in the model output.
Meanwhile, popular items are more likely to be treated as positive samples during training, which leads to higher propensity factor scores and subsequently higher interaction scores, with popularity being mistakenly interpreted as user preference.
Building on this insight, we propose a fair sampling (FS) method that ensures each user and each item has an equal likelihood of being selected as both positive and negative instances during training, thus preventing popular items from being overly selected as positive samples.
Theoretical analysis demonstrates that the proposed FS can effectively eliminate the influence of propensity factors without the need to estimate the propensity score, thereby mitigating popularity bias and overcoming the limitations of IPS-based methods.

FS is a sampling-level optimization method that can be applied to both point-wise and pair-wise loss functions.
Depending on the type of loss function, we implement FS-Point for point-wise loss and FS-Pair for pair-wise loss.
Experimental results show that FS achieves state-of-the-art performance.

\section{Related Work}\label{sec:dnm9}
IPS-based methods mitigate popularity bias in recommender systems through sample weighting.
We categorize related work by the type of loss function: point-wise or pair-wise.

\textbf{Point-wise.}
Rel-MF~\cite{saito2020unbiased} formulates interaction occurrence as a two-step process---exposure and interaction---and introduces an unbiased point-wise loss function.
DU~\cite{lee2021dual} refines this method by enhancing propensity score estimation.
CJMF~\cite{zhu2020unbiased} introduces a joint learning framework to simultaneously model unbiased user-item relevance and propensity.
BISER~\cite{lee2022bilateral} addresses item bias via self-inverse propensity weighting and employs bilateral unbiased learning to unify two complementary models.
More recently, ReCRec~\cite{lin2024recrec} improves bias correction by distinguishing unexposed from disliked items, enabling better reasoning over unclicked data.

\textbf{Pair-wise.}
UBPR~\cite{saito2020unbiased2} builds on the two-step interaction assumption to introduce an unbiased pair-wise loss function.
UPL~\cite{ren2023unbiased} extends this approach with a variance-reduced learning method.
CPR~\cite{wan2022cross} redefines unbiasedness in ranking and leverages cross pairs to improve unbiased learning.
UpliftRec~\cite{chen2024treatment} applies uplift modeling to dynamically optimize top-N recommendations, revealing latent user preferences while maximizing click-through rates.
Additionally, PU~\cite{cao2024practically} models feedback labels as a noisy proxy for exposure outcomes and integrates a theoretically noise-resistant loss function into propensity estimation.

\section{Methods}
We first introduce the notations in Section \ref{sec:notations} and formalize the ideal and classic loss functions in Sections \ref{sec:ideal} and \ref{sec:classic}, respectively.
Then, in Section \ref{sec:analysis}, we analyze the bias in the classic loss function by comparing its optimization objective with that of the ideal loss function.
Finally, in Section \ref{sec:fair}, we present the proposed fair sampling (FS) method and analyze how it mitigates popularity bias.

\subsection{Notations}\label{sec:notations}
For a user $u\in\mathcal{U}$ and an item $i\in\mathcal{I}$, we refer to a user-item pair as a \textit{positive pair} if an interaction occurs between them; otherwise, we call it a \textit{negative pair}.
A interaction occurs through a \textit{two-step process}~\cite{saito2020unbiased,saito2020unbiased2}: first, a user \textbf{observes} an item; then, based on how \textbf{relevant} the item is to the user's preferences, the user decide whether to \textbf{interact} with it.
Next, we define three binary indicator matrices:
(1) \textbf{Observation matrix} $O\in\{0,1\}^{|\mathcal{U}|\times|\mathcal{I}|}$, where $O_{u,i}$ indicates whether $u$ has observed $i$ (or, equivalently, whether $i$ has been exposed to $u$);
(2) \textbf{Relevance matrix} $R\in\{0,1\}^{|\mathcal{U}|\times|\mathcal{I}|}$, where $R_{u,i}$ indicates whether $u$ would interact with $i$ given $O_{u,i}=1$ (or, equivalently, whether $u$ likes $i$);
(3) \textbf{Interaction matrix} $Y\in\{0,1\}^{|\mathcal{U}|\times|\mathcal{I}|}$, where $Y_{u,i}$ indicates whether $u\text{-}i$ is a positive pair.
Note that only $Y$ is observable, while both $O$ and $R$ are latent variables.

Clearly, $u\text{-}i$ is a positive pair ($Y_{u,i}=1$) if and only if $u$ has observed $i$ ($O_{u,i}=1$) and the two have high relevance ($R_{u,i}=1$).
This can be expressed as $Y_{u,i}=O_{u,i}\cdot R_{u,i}$, or in matrix form, $Y=O\odot R$.
Consequently,
\begin{equation}
\begin{aligned}
P(Y_{u,i}=1)&=P(O_{u,i}=1,R_{u,i}=1) \\
&=P(O_{u,i}=1|R_{u,i}=1)\cdot P(R_{u,i}=1)\text{,}
\end{aligned}
\end{equation}
where $P(R_{u,i}=1)$ denotes the relevance probability between $u$ and $i$, and $P(O_{u,i}=1|R_{u,i}=1)$ represents the exposure probability---the probability that $i$ is observed by $u$ given their relevance.
Assuming the exposure probability can be decomposed into user propensity, item propensity, and user-item relevance, it can be expressed as:
\begin{equation}
P(O_{u,i}=1|R_{u,i}=1)=\theta_u\cdot \theta_i\cdot P(R_{u,i}=1)^\alpha\text{.}
\end{equation}
Here, $\theta_u$ and $\theta_i$ are user-specific and item-specific propensity factors, respectively, which tend to have higher values for active users and popular items.
The term $P(R_{u,i}=1)^\alpha$ indicates that a higher relevance score increases the exposure probability, where $\alpha$ is a positive constant used to moderate its influence.

\subsection{Ideal Point-wise Loss and Pair-wise Loss}\label{sec:ideal}
Ideally, a recommendation model should reflect users' intrinsic preferences $R$ rather than simply imitating observed user behaviors $Y$.
For this to hold, users must observe \textbf{all} items and decide whether to interact with them based on relevance.
In this scenario, all elements of $O$ are assigned a value of $1$, making $Y$ identical to $R$.

\subsubsection{Ideal Point-wise Loss}
One commonly used point-wise loss is the cross-entropy loss.
Let $s_{u,i}$ denote the predicted score of $u\text{-}i$ produced by the model, then $\delta^+(u,i) = -\log(s_{u,i})$ and $\delta^-(u,i) = -\log(1 - s_{u,i})$, where $\delta^+(u,i)$ and $\delta^-(u,i)$ denote the loss contributions for $u\text{-}i$ being a positive and a negative pair, respectively.
The \textbf{ideal point-wise loss} is defined as follows:
\begin{equation}\label{eq:vj2j}
\mathcal{L}_{\text{ideal}}^{\text{point}}=\sum_{(u,i;R_{u,i})\in\mathcal{U}\times\mathcal{I}}\left[R_{u,i}\cdot\delta^+(u,i)+(1-R_{u,i})\cdot\delta^-(u,i)\right]\text{.}
\end{equation}

\subsubsection{Ideal Pair-wise Loss}
One commonly used pair-wise loss is the BPR loss, which is defined as $\zeta(u,i,j)=-\ln{\sigma{(s_{u,i}-s_{u,j})}}$, where $\sigma(\cdot)$ is the sigmoid function.
The \textbf{ideal pair-wise loss} is defined as follows:
\begin{equation}\label{eq:h92m}
\mathcal{L}_{\text{ideal}}^{\text{pair}}=\sum_{(u,i,j;R_{u,i},R_{u,j})\in\mathcal{U}\times\mathcal{I}\times{\mathcal{I}}}{\left[R_{u,i}\cdot(1-R_{u,j})\cdot\zeta(u,i,j)\right]}\text{.}
\end{equation}

\subsubsection{Discussion}
The ideal loss functions can guide the model to learn users' inherent preferences.
However, since $R$ is unobservable, the ideal loss functions \textbf{cannot} be computed directly.

\subsection {Classic Point-wise Loss and Pair-wise Loss}\label{sec:classic}

\subsubsection{Classic Point-wise Loss}
WMF~\cite{hu2008collaborative} introduced a \textbf{point-wise loss} for modeling implicit feedback data:
\begin{equation}\label{eq:cm9d}
\mathcal{L}_{\text{bias}}^{\text{point}}=\sum_{(u,i;Y_{u,i})\in\mathcal{U}\times\mathcal{I}}\left[Y_{u,i}\cdot\delta^+(u,i)+(1-Y_{u,i})\cdot\delta^-(u,i)\right]\text{.}
\end{equation}

\subsubsection{Classic Pair-wise Loss}
BPR~\cite{rendle2012bpr} introduced a \textbf{pair-wise loss} for modeling implicit feedback data:
\begin{equation}\label{eq:19mc}
\mathcal{L}_{\text{bias}}^{\text{pair}}=\sum_{(u,i,j;Y_{u,i},Y_{u,j})\in\mathcal{U}\times\mathcal{I}\times{\mathcal{I}}}{\left[Y_{u,i}\cdot(1-Y_{u,j})\cdot\zeta(u,i,j)\right]}\text{.}
\end{equation}

\subsubsection{Discussion}
Classic loss functions are designed to directly model the interaction matrix $Y$.
Since $Y$ depends not only on $R$ but also on $O$, classic loss functions provide biased approximations of the ideal loss functions.

\subsection{Analysis}\label{sec:analysis}
Let $s_{u,i}^{\text{ideal}}$ and $s_{u,i}^{\text{bias}}$ denote the predicted results obtained by optimizing the model with the ideal loss ($\mathcal{L}_{\text{ideal}}^{\text{point}}$ or $\mathcal{L}_{\text{ideal}}^{\text{pair}}$) and the classic loss ($\mathcal{L}_{\text{bias}}^{\text{point}}$ or $\mathcal{L}_{\text{bias}}^{\text{pair}}$), respectively.
$s_{u,i}^{\text{ideal}}$ captures users' intrinsic preferences $R$, while $s_{u,i}^{\text{bias}}$ reflects users' interaction behaviors $Y$.
Therefore,
\begin{equation}\label{eq:cvna}
\begin{aligned}
s_{u,i}^{\text{ideal}}&=P(R_{u,i}=1)>0\text{,} \\
s_{u,i}^{\text{bias}}&=P(Y_{u,i}=1)=\theta_{u}\theta_i P(R_{u,i}=1)^{\alpha+1}>0\text{.}
\end{aligned}
\end{equation}

Intuitively, classic loss functions introduce biases as they inherently incorporate propensity factors into the model's output.
As a result, even if user $u$ exhibits the same level of preference for both items $i$ and $j$ ($P(R_{u,i}=1)=P(R_{u,j}=1)$), the optimization outcome of classic loss functions tends to favor the item with a higher propensity factor ($\theta_{u}\theta_{i}P(R_{u,i}=1)^\alpha$ or $\theta_{u}\theta_{j}P(R_{u,j}=1)^\alpha$).

\subsubsection{Point-wise Loss}
The optimization objective of $\mathcal{L}_{\text{ideal}}^{\text{point}}$ is to increase $s_{u,i}^{\text{ideal}}$ when $R_{u,i} = 1$ and decrease $s_{u,i}^{\text{ideal}}$ when $R_{u,i} = 0$.
We use $\Delta x$ to denote the change in $x$ after one optimization step (e.g., gradient descent).
Therefore,
\begin{equation}
\begin{cases}  
\Delta{s_{u,i}^{\text{ideal}}}=\Delta{P(R_{u,i}=1)}>0 & \text{if } R_{u,i}=1\text{,} \\  
-\Delta{s_{u,i}^{\text{ideal}}}=-\Delta{P(R_{u,i}=1)}>0 & \text{if } R_{u,i}=0\text{,}
\end{cases}
\end{equation}
which can also be written as:
\begin{equation}\label{eq:bmad}
\begin{cases}  
\Delta{\ln{P(R_{u,i}=1)}}>0 & \text{if } R_{u,i}=1\text{,} \\  
-\Delta{\ln{P(R_{u,i}=1)}}>0 & \text{if } R_{u,i}=0\text{.}
\end{cases}
\end{equation}
While for $\mathcal{L}_{\text{bias}}^{\text{point}}$, its optimization objective is to increase $s_{u,i}^{\text{bias}}$ when $Y_{u,i} = 1$ and decrease $s_{u,i}^{\text{bias}}$ when $Y_{u,i} = 0$.
Therefore,
\begin{equation}
\begin{cases}  
\Delta{s_{u,i}^{\text{bias}}}=\Delta{P(Y_{u,i}=1)}>0 & \text{if } Y_{u,i}=1\text{,} \\  
-\Delta{s_{u,i}^{\text{bias}}}=-\Delta{P(Y_{u,i}=1)}>0 & \text{if } Y_{u,i}=0\text{,}
\end{cases}
\end{equation}
which can also be written as\footnote{Note that $\Delta(x+y)=\Delta x+\Delta y$, and $\Delta ax=a\Delta x$, where $a$ is a constant.}:
\begin{equation}\label{eq:v8e2}
\begin{cases}
\Delta{\ln{P(Y_{u,i}=1)}}\\=\Delta{\ln{\theta_u}}+\Delta{\ln{\theta_i}}+(\alpha+1)\Delta{\ln{P(R_{u,i}=1)}}>0 & \text{if } Y_{u,i}=1\text{,} \\  
-\Delta{\ln{P(Y_{u,i}=1)}}\\=-\Delta{\ln{\theta_u}}-\Delta{\ln{\theta_i}}-(\alpha+1)\Delta{\ln{P(R_{u,i}=1)}}>0 & \text{if } Y_{u,i}=0\text{.}
\end{cases}
\end{equation}

\subsubsection{Pair-wise Loss}
Assume that $u\text{-}i$ is a positive pair, and $u\text{-}j$ is a negative pair.
The optimization objective of $\mathcal{L}_{\text{ideal}}^{\text{pair}}$ is to increase the margin between $s_{u,i}^{\text{ideal}}$ and $s_{u,j}^{\text{ideal}}$.
Therefore,
\begin{equation}\label{eq:948f}
\Delta{(\ln{P(R_{u,i}=1)}-\ln{P(R_{u,j}=1))}}=\Delta{\ln{\frac{P(R_{u,i}=1)}{P(R_{u,j}=1)}}}>0\text{.}
\end{equation}
While for $\mathcal{L}_{\text{bias}}^{\text{pair}}$, its optimization objective it to increase the margin between $s_{u,i}^{\text{bias}}$ and $s_{u,j}^{\text{bias}}$.
Therefore,
\begin{equation}\label{eq:v932}
\begin{aligned}
&\quad\Delta{(\ln{P(Y_{u,i}=1)}-\ln{P(Y_{u,j}=1))}} \\
&=\Delta{\ln{\frac{\theta_i}{\theta_j}}}+(\alpha+1)\Delta{\ln{\frac{P(R_{u,i}=1)}{P(R_{u,j}=1)}}}>0\text{.}
\end{aligned}
\end{equation}

\subsubsection{Discussion}
By comparing Equation (\ref{eq:bmad}) with (\ref{eq:v8e2}), and Equation (\ref{eq:948f}) with (\ref{eq:v932}), it can be observed that $\mathcal{L}_{\text{bias}}^{\text{point}}$ and $\mathcal{L}_{\text{bias}}^{\text{pair}}$ improperly optimize the propensity factors ($\theta_u$ and $\theta_i$ for $\mathcal{L}_{\text{bias}}^{\text{point}}$, and $\theta_i$ and $\theta_j$ for $\mathcal{L}_{\text{bias}}^{\text{pair}}$).
Since popular items and active users are more frequently selected to form positive pairs during training, their propensity factors tend to be higher.
Ultimately, propensity factors related to exposure probability $P(O_{u,i}=1|R_{u,i}=1)$, rather than relevance probability $P(R_{u,i}=1)$, lead to higher interaction scores $P(Y_{u,i}=1)$, which is not the intended outcome.


\begin{table*}\footnotesize
  \caption{Overall performance comparison.}
  \label{tab:956d}
\begin{tabular}{ccc|ccc|ccc|ccc}
\hline
\multicolumn{3}{c|}{\multirow{2}{*}{Model}}                                                                          & \multicolumn{3}{c|}{Kindle}                       & \multicolumn{3}{c|}{Gowalla}                      & \multicolumn{3}{c}{Yelp}                          \\ \cline{4-12} 
\multicolumn{3}{c|}{}                                                                                                & Recall $\uparrow$          & NDCG $\uparrow$            & ARP $\downarrow$           & Recall $\uparrow$          & NDCG $\uparrow$            & ARP $\downarrow$           & Recall $\uparrow$          & NDCG $\uparrow$            & ARP $\downarrow$           \\ \hline
\multicolumn{1}{c|}{\multirow{6}{*}{Point-wise}} & \multicolumn{1}{c|}{\multirow{3}{*}{IPS}}    & WMF               & 0.1095          & 0.0979          & 1896          & 0.1052          & 0.0924          & 3995          & 0.0398          & 0.0297          & 4231          \\
\multicolumn{1}{c|}{}                             & \multicolumn{1}{c|}{}                        & Rel-MF            & 0.1192          & 0.1088          & 1660          & 0.1116          & 0.0975          & 3800          & 0.0471          & 0.0359          & 3651          \\
\multicolumn{1}{c|}{}                             & \multicolumn{1}{c|}{}                        & DU                & 0.1264          & 0.1141          & 1497          & 0.1168          & 0.1018          & 3356          & 0.0504          & 0.0382          & 3286          \\ \cline{2-12} 
\multicolumn{1}{c|}{}                             & \multicolumn{1}{c|}{\multirow{2}{*}{Causal}} & ExpoMF            & 0.1353          & 0.1226          & 1354          & 0.1186          & 0.1013          & 3173          & 0.0514          & 0.0401          & 3105          \\
\multicolumn{1}{c|}{}                             & \multicolumn{1}{c|}{}                        & CauseE            & 0.1336          & 0.1233          & 1447          & 0.1179          & 0.1005          & 3159          & 0.0528          & 0.0408          & 3025          \\ \cline{2-12} 
\multicolumn{1}{c|}{}                             & \multicolumn{1}{c|}{\textbf{Our}}            & \textbf{FS-Point} & \textbf{0.1413} & \textbf{0.1263} & \textbf{1175} & \textbf{0.1217} & \textbf{0.1067} & \textbf{2261} & \textbf{0.0554} & \textbf{0.0437} & \textbf{2420} \\ \hline
\multicolumn{1}{c|}{\multirow{6}{*}{Pair-wise}}  & \multicolumn{1}{c|}{\multirow{3}{*}{IPS}}    & BPR               & 0.1431          & 0.1169          & 1572          & 0.1179          & 0.0943          & 3703          & 0.0527          & 0.0388          & 3939          \\
\multicolumn{1}{c|}{}                             & \multicolumn{1}{c|}{}                        & UBPR              & 0.1414          & 0.1196          & 1156          & 0.1125          & 0.0936          & 2447          & 0.0600          & 0.0452          & 2489          \\
\multicolumn{1}{c|}{}                             & \multicolumn{1}{c|}{}                        & CPR               & 0.1516          & 0.1275          & 1406          & 0.1240          & 0.1023          & 2509          & 0.0634          & 0.0478          & 2901          \\ \cline{2-12} 
\multicolumn{1}{c|}{}                             & \multicolumn{1}{c|}{\multirow{2}{*}{Causal}} & DICE              & 0.1454          & 0.1195          & 1303          & 0.1160          & 0.0966          & 2302          & 0.0599          & 0.0455          & 2459          \\
\multicolumn{1}{c|}{}                             & \multicolumn{1}{c|}{}                        & PD                & 0.1418          & 0.1214          & 1254          & 0.1162          & 0.0953          & 2388          & 0.0613          & 0.0469          & 2404          \\ \cline{2-12} 
\multicolumn{1}{c|}{}                             & \multicolumn{1}{c|}{\textbf{Our}}            & \textbf{FS-Pair}  & \textbf{0.1652} & \textbf{0.1384} & \textbf{865}  & \textbf{0.1294} & \textbf{0.1052} & \textbf{1140} & \textbf{0.0741} & \textbf{0.0552} & \textbf{1534} \\ \hline
\end{tabular}
\end{table*}

\subsection{Fair Sampling}\label{sec:fair}
Fair sampling (FS) constructs supplementary sample(s) for each original sample in the classic loss during model training, helping to prevent the improper optimization of propensity factors and thereby mitigating popularity bias.
Depending on the type of loss function, FS has two variants: \textbf{FS-Point}, designed for point-wise loss, and \textbf{FS-Pair}, designed for pair-wise loss.

\subsubsection{FS-Point}
The sample set utilized by the classic point-wise loss function $\mathcal{L}_{\text{bias}}^{\text{point}}$ is:
\begin{equation}
\mathcal{D}^{\text{point}}_\text{bias}=\{(u,i;Y_{u,i})|u\in\mathcal{U},i\in\mathcal{I}\}\text{.}
\end{equation}
For each $(u,i;Y_{u,i})\in\mathcal{D}^{\text{point}}_\text{bias}$, we can find a corresponding $(\tilde{u},\tilde{i})$ that satisfies the conditions $Y_{\tilde{u},\tilde{i}}=Y_{u,i}$, $Y_{\tilde{u},i}=1-Y_{u,i}$, and $Y_{u,\tilde{i}}=1-Y_{u,i}$.
Then, we can obtain the sample set for FS-Point loss by combining them together:
\begin{equation}
\mathcal{D}^{\text{point}}_\text{FS}=\mathcal{D}^{\text{point}}_\text{bias} \cup \{(\tilde{u},\tilde{i};Y_{\tilde{u},\tilde{i}}),(\tilde{u},i;Y_{\tilde{u},i}),(u,\tilde{i};Y_{u,\tilde{i}})|(u,i;Y_{u,i})\in\mathcal{D}^{\text{point}}_\text{bias}\}\text{.}
\end{equation}
Finally, \textbf{FS-Point loss} is defined as:
\begin{equation}
\mathcal{L}_{\text{FS}}^{\text{point}}=\sum_{(u,i;Y_{u,i})\in\mathcal{D}^{\text{point}}_\text{FS}}\left[Y_{u,i}\cdot\delta^+(u,i)+(1-Y_{u,i})\cdot\delta^-(u,i)\right]\text{.}
\end{equation}

We assume $Y_{u,i}=1$ (a similar conclusion can be drawn when $Y_{u,i}=0$).
Based on Equation (\ref{eq:v8e2}), when $(u,i;Y_{u,i}=1)$ is input into $\mathcal{L}_{\text{FS}}^{\text{point}}$ for optimization, $\ln\theta_{u}$ and $\ln\theta_i$ are amplified.
However, when the corresponding supplementary samples---$(\tilde{u},\tilde{i};Y_{\tilde{u},\tilde{i}}=1)$, $(\tilde{u},i;Y_{\tilde{u},i}=0)$, and $(u,\tilde{i};Y_{u,\tilde{i}}=0)$---are input into $\mathcal{L}_{\text{FS}}^{\text{point}}$ for optimization, the optimization effect on the propensity factors is offset:
\begin{equation}\label{eq:9gm4}
\begin{aligned}
&(\Delta{\ln\theta_u}+\Delta{\ln\theta_i})+(\Delta{\ln\theta_{\tilde{u}}}+\Delta{\ln\theta_{\tilde{i}}}) + \\
&(-\Delta{\ln\theta_{\tilde{u}}}-\Delta{\ln\theta_i})+(-\Delta{\ln\theta_u}-\Delta{\ln\theta_{\tilde{i}}})=0\text{.}
\end{aligned}
\end{equation}

\subsubsection{FS-Pair}
The sample set utilized by the classic pair-wise loss function $\mathcal{L}_{\text{bias}}^{\text{pair}}$ is:
\begin{equation}
\mathcal{D}_\text{bias}^{\text{pair}}=\{(u,i,j;Y_{u,i},Y_{u,j})|u\in\mathcal{U},i,j\in\mathcal{I}\}\text{.}
\end{equation}
For each $(u,i,j;Y_{u,i},Y_{u,j})\in\mathcal{D}_\text{bias}^{\text{pair}}$, we can find a corresponding $(\tilde{u},j,i)$ that satisfies the conditions $Y_{\tilde{u},j}=Y_{u,i}$ and $Y_{\tilde{u},i}=Y_{u,j}$.
Then, we can obtain the sample set for FS-Pair loss by combining them together:
\begin{equation}
\mathcal{D}_\text{FS}^{\text{pair}}=\mathcal{D}_\text{bias}^{\text{pair}}\cup\{(\tilde{u},j,i;Y_{\tilde{u},j},Y_{\tilde{u},i})|(u,i,j)\in\mathcal{D}_\text{bias}^{\text{pair}}\}
\end{equation}
Finally, \textbf{FS-Pair loss} is defined as:
\begin{equation}
\mathcal{L}_{\text{FS}}^{\text{pair}}=\sum_{(u,i,j;Y_{u,i},Y_{u,j})\in\mathcal{D}_\text{FS}^{\text{pair}}}{\left[Y_{u,i}\cdot(1-Y_{u,j})\cdot\zeta(u,i,j)\right]}\text{.}
\end{equation}

Since a sample contributes to the loss only when $Y_{u,i}=1$ and $Y_{u,j}=0$, we focus on this case.
Based on Equation (\ref{eq:v932}), when $(u,i,j;Y_{u,i}=1,Y_{u,j}=0)$ is input into $\mathcal{L}_{\text{FS}}^{\text{pair}}$ for optimization, $\ln{\frac{\theta_i}{\theta_j}}$ is amplified.
However, when the corresponding supplementary sample $(\tilde{u},j,i;Y_{\tilde{u},j}=1,Y_{\tilde{u},i}=0)$ is input into $\mathcal{L}_{\text{FS}}^{\text{pair}}$ for optimization, $\ln{\frac{\theta_j}{\theta_i}}$ is amplified, which offsets the optimization effect on the propensity factors:
\begin{equation}\label{eq:ib93}
\Delta{\ln{\frac{\theta_i}{\theta_j}}}+\Delta{\ln{\frac{\theta_j}{\theta_i}}}=\Delta{(\ln{\frac{\theta_i}{\theta_j}}+\ln{\frac{\theta_j}{\theta_i}})}=0\text{.}
\end{equation}

\subsubsection{Discussion}
The idea of FS is simple---whenever a user or item is chosen to form a positive/negative pair, the user or item is simultaneously selected to form a corresponding negative/positive pair.
Equations (\ref{eq:9gm4}) and (\ref{eq:ib93}) indicate that FS-Point loss and FS-Pair loss no longer optimize propensity factors.
Consequently, only the preference-related factor $P(R_{u,i}=1)$ is optimized, effectively mitigating the popularity bias.
Note that FS does not require estimating the propensity score, which avoids the incomplete removal of popularity bias in IPS-based methods, which stems from their inability to estimate the propensity score accurately.

\section{Experiments}

\subsection{Settings}

\subsubsection{Datasets}
We experiment with three widely used datasets: Amazon Review (Kindle)~\cite{ni2019justifying}, Gowalla~\cite{cho2011friendship}, and Yelp.
These datasets contain log data capturing observed user behavior.
Under the two-step interaction assumption, these datasets exhibit popularity bias.
We retain only interactions with ratings $\ge 4$ and ensure that both users and items have at least 20 interactions.

Following prior offline evaluation protocols~\cite{bonner2018causal,liang2016causal,wu2022discovering}, we construct unbiased validation and test sets by sampling from the full dataset with equal selection probability for each item.
The remaining data serves as the training set.
The dataset is split into training, validation, and test sets in a 7:1:2 ratio.

\subsubsection{Evaluation}
We evaluate performance using three metrics: Recall@K, NDCG@K, and ARP@K, with $K$ set to 20 by default.
Average Recommendation Popularity at $K$ (ARP@K) serves as a complementary metric for measuring recommendation bias~\cite{yin2012challenging,abdollahpouri2019managing}.
It calculates the average popularity of the top-$K$ recommended items per user, where lower ARP@K values indicate reduced bias.

\subsubsection{Baselines}
In addition to the IPS-based methods, we also use four causal inference methods---ExpoMF~\cite{liang2016modeling}, CauseE~\cite{bonner2018causal}, PD~\cite{zhang2021causal}, and DICE~\cite{zheng2021disentangling}---as baselines for comparison.
To ensure fairness, all methods use MF~\cite{koren2009matrix} as the backbone model and share the same hyperparameter search space.
All point-wise methods use cross-entropy loss, while all pair-wise methods use BPR loss.

\subsection{Results}
As shown in Table~\ref{tab:956d}, FS-Point achieves the highest Recall and NDCG, and the lowest APR among point-wise learning methods.
Similarly, FS-Pair achieves the best performance in these metrics among pair-wise methods.
This suggests that, compared to other baselines, FS more effectively mitigates the influence of popularity bias, increasing the likelihood of recommending unpopular items without sacrificing recommendation accuracy.
Notably, despite its simplicity, theoretical analysis confirms that FS completely eliminates popularity bias, enabling optimal performance under the current evaluation strategy, where the test set is designed to be unaffected by item popularity.

\section{Conclusions}
We propose a fair sampling (FS) method, which mitigates popularity bias in collaborative filtering by ensuring that each item appears equally as both a positive and negative sample during training.
Both theoretical analysis and experimental results demonstrate the effectiveness of the proposed FS method. 
A potential limitation of FS is that it completely overlooks the influence of popularity, even though popularity is often correlated with higher quality.
Strategically adjusting the sampling ratio can help balance the diversity and quality of recommendations.

\begin{acks}
This work is supported by National Natural Science Foundation of China (NSFC) under the Grant No. 62372113. Peng Zhang is a faculty of School of Computer Science, Fudan University. Tun Lu is a faculty of School of Computer Science, Shanghai Key Laboratory of Data Science, Fudan Institute on Aging, MOE Laboratory for National Development and Intelligent Governance, and Shanghai Institute of Intelligent Electronics \& Systems, Fudan University.
\end{acks}

\bibliographystyle{ACM-Reference-Format}
\balance
\bibliography{main}

\appendix

\end{document}